\documentclass[aip,pop,reprint,superscriptaddress,preprintnumbers]{revtex4-1}
\usepackage{bm}
\usepackage[colorlinks=true,allcolors=blue]{hyperref}
\usepackage{graphicx}
\usepackage{color}

\begin{document}

\preprint{LLNL-JRNL-759681-DRAFT}

\title{Full-$f$ gyrokinetic simulation of turbulence in a helical open-field-line plasma}


\author{E.\,L. Shi}
\affiliation{Lawrence Livermore National Laboratory, Livermore, California 94550, USA}
\affiliation{Department of Astrophysical Sciences, Princeton University, Princeton, New Jersey 08544, USA}

\author{G.\,W. Hammett}
\email[]{hammett@pppl.gov}
\affiliation{Princeton Plasma Physics Laboratory, Princeton, New Jersey 08540, USA}

\author{T. Stoltzfus-Dueck}
\affiliation{Princeton Plasma Physics Laboratory, Princeton, New Jersey 08540, USA}

\author{A. Hakim}
\affiliation{Princeton Plasma Physics Laboratory, Princeton, New Jersey 08540, USA}


\date{\today}

\begin{abstract}
Curvature-driven turbulence in a helical open-field-line plasma is investigated using electrostatic
five-dimensional gyrokinetic continuum simulations in an all-bad-curvature
    helical-slab geometry.
Parameters for a National Spherical Torus Experiment scrape-off-layer plasma are used in the model.
The formation and convective radial transport of plasma blobs is observed, and it is shown that
the radial particle-transport levels are several times higher than diffusive
    Bohm-transport estimates.
By reducing the strength of the poloidal magnetic field, the profile of the
heat flux to the divertor plate is observed to broaden.
\end{abstract}


\maketitle

\section{Introduction}
Satisfactory and reliable quantitative predictions of turbulence and transport in the
tokamak edge and scrape-off-layer (SOL) regions are widely believed
to require the use of expensive gyrokinetic simulations in some capacity.\cite{Cohen2008,Ricci2015b,Scott2010b,Scott2003}
Some major outstanding questions that require numerical investigation include
how the SOL power width is set, \cite{Eich2013,Goldston2012,Chang2017b}
how a confined plasma transitions from a low-confinement $L$ mode to a
high-confinement $H$ mode,\cite{Wagner2007,Wagner1982} and how high the
$H$-mode pedestal temperature can get, since the pedestal temperature has a
major impact on the core temperature profile and the resulting gain.
\cite{Kotschenreuther1995,Kinsey2011}
Gyrokinetic simulations in the edge and SOL regions are challenging for
several reasons (e.g. large-amplitude fluctuations, steep profile gradients,
closed and open magnetic field lines, X-point, effective sheath-model boundary
conditions),
but specialized particle-in-cell \cite{Churchill2017,Korpilo2016} and continuum gyrokinetic codes
\cite{Shi2017,ShiPhDThesis2017,Pan2018,Dorf2016,Pan2016} have been making
steady progress towards
the ultimate goal as a predictive tool for boundary-plasma modeling.
We refer the reader to \citet{Cohen2008} for a summary of early approaches to
gyrokinetic edge and SOL simulation and \citet{Krommes2012} for an
introduction to gyrokinetics.
The particle-in-cell-based XGC1 code\cite{Chang2009} is the only
gyrokinetic code at present
that is able to simulate turbulence in a three-dimensional diverted geometry.
The first gyrokinetic simulations using continuum algorithms to simulate
turbulence on straight open field lines were presented in \citet{Shi2017}
and then in \citet{Pan2018}.
Here and in \citet{ShiPhDThesis2017}, we extend this earlier work to present
the first gyrokinetic continuum simulations on open field lines including
curved toroidal fields, which can strongly enhance the drive of plasma
instabilities.

The SOL refers to the tokamak plasma region of open magnetic field lines between
the last closed flux surface (LCFS) and the first wall.
Here, the field lines intersect material surfaces that act as plasma sinks where the loss rate of electrons and ions
are kept in approximate particle balance by a Debye sheath layer.
Plasma--surface interactions \cite{Stangeby2000} at the material interfaces can also
contaminate the plasma with wall materials, which can severely
degrade the fusion-plasma quality,
but we do not yet incorporate these effects in the model being presented here.

Probe and imaging diagnostics have revealed the existence of
intermittent coherent structures in the SOL referred to as \textit{plasma filaments} or \textit{blobs},
\citep{Zweben2004,Terry2007,Boedo2014,Zweben1985b,Zweben1985a}
which convectively transport particles, heat, momentum, and current across magnetic field lines.\citep{DIppolito2011}
Blobs are characterized by densities that are much higher than local background levels,
a structure that is highly elongated along the magnetic field (much larger than
the plasma minor radius),
and much smaller scales perpendicular to the magnetic field, ${\sim}10 \rho_i$,
where $\rho_i$ is the ion gyroradius.\citep{DIppolito2011,Zweben2007}
Cross-field transport in the far SOL is highly intermittent due to blob
propagation \citep{Zweben2007} and
is consequently poorly described in terms of effective diffusion coefficients and
convective velocities.\citep{Naulin2007}

In a tokamak, the curvature and $\nabla \bm{B}$ forces are believed to set
up a charge-separated dipole potential structure across the blob
cross-section that results in its outward radial propagation via
convective $\bm{E}\times\bm{B}$ transport.\cite{Krasheninnikov2001,DIppolito2011}
Finite-temperature effects of the blob can also cause spin motion if the
blob is sheath-connected, which can reduce this radial motion.\cite{Myra2004}
Numerically, blobs dynamics have been studied using seeded-blob fluid
simulations.\cite{Angus2012,Riva2016,Walkden2015,Shanahan2016}
Self-consistent blob formation has been studied with two-dimensional models
\cite{Bisai2005,Sarazin2003,Garcia2006} and in three-dimensional
turbulence simulations.\cite{Churchill2017,Ricci2013,Stegmeir2018,Baudoin2018}

The work presented here builds on our previous efforts in simulating open-magnetic-field-line turbulence
in the Large Plasma Device \cite{Gekelman2016}
using the gyrokinetic continuum capabilities of the Gkeyll code.\cite{Shi2017}
In that study, the magnetic field was straight and uniform, and the plasma was highly collisional,
which necessitated the use of an artificial electron-to-ion mass ratio ($m_i/m_e = 400$)
and reduced electron collision frequencies to make the simulations tractable,
given the explicit algorithm used at present for the collision operator.
Nevertheless, we found that our numerical approach based on discontinuous Galerkin methods
and sheath-model boundary conditions for an open-field-line region
were stable and produced qualitatively reasonable results,
which led to the first demonstration of open-field-line turbulence with a gyrokinetic continuum code.
The reduced electron mass and collision-frequency restrictions have been
relaxed for the simulations presented in this paper,
which now also include a more sophisticated magnetic geometry.

We have added magnetic curvature and $\nabla \bm{B}$ drifts to the Gkeyll code
and can simulate
a helical magnetic geometry approximating that in simple magnetized tori
(SMT's), such as TORPEX \cite{Fasoli2006} and Helimak.\cite{Gentle2008}
In contrast to early work on the simulation of turbulence in SMT's based
on the drift-reduced Braginskii equations and neglecting the ion temperature,
\cite{Ricci2008,Ricci2009a,Li2011}
the gyrokinetic approach naturally can investigate plasmas with
$T_i \gtrsim T_e$, which is commonly observed in the SOL.\cite{Boedo2009,Kocan2011,Kocan2012}
Recent fluid simulations are also including finite $T_i$.
\cite{Halpern2016,Zhu2017}

Although our simulations do not yet simultaneously contain open- and
closed-field-line regions,\cite{Ribeiro2005,Zweben2009,
Halpern2016,Dudson2017,Zhu2017,Francisquez2017}
we believe that many basic properties of SOL turbulence and transport are reproduced in this model.
Additionally, the turbulence in this helical open-field-line geometry with parameters appropriate for 
a tokamak SOL has not been previously studied using a gyrokinetic PIC approach, either.
We do acknowledge, however, that gyrokinetic PIC codes that have the necessary capabilities for
the problem described in this paper have already been developed,\cite{Churchill2017,Korpilo2016}
and it should be straightforward for these codes to implement this simple
helical geometry for cross-code comparisons.

We discuss details of the helical-SOL model in Sec.~\ref{sec:model},
including equations solved, simulation geometry, and boundary conditions.
Additional details about the underlying algorithms can be found in
Refs.~\onlinecite{Shi2017} and \onlinecite{ShiPhDThesis2017}.
We present simulation results obtained using the Gkeyll code in Sec.~\ref{sec:results},
such as heat-flux profiles, fluctuation statistics, and particle fluxes.
Our conclusions are given in Sec.~\ref{sec:conclusions}.
To facilitate future code comparisons, we also present details of the
initial conditions used in our simulations in Appendix~\ref{sec:initial-helical-sol}.

\section{Model \label{sec:model}}
In the non-orthogonal field-aligned geometry used in the simulation,
$z$ measures distances along field lines relative to the midplane (poloidal
angle $\theta_\mathrm{pol}=0$ in a tokamak), $x$ is the radial coordinate, and $y$ is
constant along a field line and measures distances perpendicular to field
lines.
The simulation geometry is a flux tube on the outboard side that wraps
around the torus a specified number of times, terminating on material 
surfaces at each end in $z$.
The resulting mapping from field-aligned coordinates $(x,y,z)$ to standard
cylindrical coordinates $(R,\varphi, Z)$ is given by $R=x$, $Z=z \sin \theta$,
and $\varphi = \left(y \sin \theta + z \cos \theta \right) / R_c$,
where $R_c=R_0 + a$ and the field-line pitch $\sin \theta = B_v / B$
are taken to be constant, $R_0$ is the device major radius, $a$ is the device
minor radius, and $B_v$ is the vertical (or poloidal) magnetic field.
This simple helical geometry has vertical flux surfaces (i.e., ignores flux
expansion) and has no magnetic shear, and some further approximations to
differential operators are made assuming short-wavelength turbulence for now.
The final model nevertheless includes the main effect of the bad-curvature
drive by toroidal magnetic fields while using an efficient 
field-aligned grid.
See Refs.~\onlinecite{Beer1995,Hammett1993,Scott1998} and
\onlinecite{ShiPhDThesis2017} for further details.

We solve a full-$f$ gyrokinetic equation written in the conservative form \cite{Brizard2007,Sugama2000,Idomura2009}
\begin{eqnarray}
  \frac{\partial \mathcal{J} f_s}{\partial t} + \nabla \! \cdot \! (\mathcal{J} \dot{\bm{R}} f_s) +
  \frac{\partial}{\partial v_\parallel}(\mathcal{J} \dot{v}_\parallel f_s) =&& \mathcal{J} C[f_s] + \mathcal{J} S_s, \label{eq:gke}
\end{eqnarray}
where $f_s = f_s(\bm{R}, v_\parallel, \mu, t)$ is the gyrocenter distribution function for species $s$,
$\mathcal{J} = B_\parallel^*$ is the Jacobian of the gyrocenter coordinates,
$B_\parallel^* = \bm{b} \cdot \bm{B^*}$, $\bm{B^*} = \bm{B} + (B v_\parallel/\Omega_s) \nabla \times \bm{b}$,
$C[f_s]$ represents the effects of collisions, $\Omega_s = q_s B/m_s$, and
$S_s = S_s(\bm{R}, v_\parallel,\mu, t)$ represents plasma sources.
The characteristics are calculated as $\dot{\bm{R}} = \{\bm{R}, H\}$ and  $\dot{v}_\parallel = \{v_\parallel, H\}$,
where the gyrokinetic Poisson bracket operator is
\begin{equation}
\{F,G\} = \frac{\bm{B^*}}{m_s B_\parallel^*} \cdot \left( \nabla F \frac{\partial G}{\partial v_\parallel} - \frac{\partial F}{\partial v_\parallel} \nabla G \right)
- \frac{1}{q_s B_\parallel^*} \bm{b} \cdot \nabla F \times \nabla G,
\end{equation}
and the gyrocenter Hamiltonian is $H_s = \frac{1}{2} m_s v_\parallel^2 + \mu B + q_s \phi$,
where the long-wavelength limit has been taken to neglect gyroaveraging.
A conservative Lenard--Bernstein collision operator \cite{Lenard1958} that 
neglects the velocity dependence of the collision frequency is used to 
model self-species and electron--ion collisions.

This system is closed by the long-wavelength gyrokinetic Poisson equation with a linearized ion polarization density
\begin{equation}
-\nabla_\perp \cdot \left( \frac{n_{i0}^g q_i^2 \rho_{\mathrm{s}0}^2}{T_{e0} } \nabla_\perp \phi \right) =  \sigma_g = q_i n_i^g(\boldsymbol{R}) - e n_e(\boldsymbol{R}), \label{eq:gkp}
\end{equation}
where $\rho_{\mathrm{s}0} = c_{\mathrm{s}0} / \Omega_i$, $c_{\mathrm{s}0} = \sqrt{T_{e0}/m_i}$, and $n_{i0}^g$ is the background
ion gyrocenter density that we take to be a constant in space and in time.

In these equations, we neglect geometrical factors arising from a cylindrical coordinate system everywhere except
in $\boldsymbol{B}^* = \boldsymbol{B} + (B v_\parallel/\Omega_s) \nabla \times \boldsymbol{b}$,
where we make the approximation that perpendicular gradients are much stronger than
parallel gradients:
\begin{eqnarray}
  (\nabla \times \boldsymbol{b}) \cdot \nabla f(x,y,z) =& \left[ (\nabla \times \boldsymbol{b}) \cdot \nabla y \right] \frac{\partial f(x,y,z)}{\partial y} \nonumber \\*
  & + \left[ (\nabla \times \boldsymbol{b}) \cdot \nabla z \right] \frac{\partial f(x,y,z)}{\partial z} \nonumber \\*
  \approx & \left[ (\nabla \times \boldsymbol{b}) \cdot \boldsymbol{e}^y \right] \frac{\partial f(x,y,z)}{\partial y}.
\end{eqnarray}
Here, we assume that $(\nabla \times \boldsymbol{b}) \cdot \boldsymbol{e}^y = -1/x$, where $\boldsymbol{e}^y = \nabla y$
is a `co-basis' direction.
This type of approximation has also been employed in some fluid simulations of
SMT's.\citep{Ricci2009a,Ricci2010}
We assume that $\boldsymbol{B} = B_\mathrm{axis} (R_0 / x) \boldsymbol{e}_z$. 

Periodic boundary conditions are applied to both $f$ and $\phi$ in $y$,
and the Dirichlet boundary condition $\phi = 0$ is applied in $x$, which
prevents gyrocenters from crossing the surfaces in $x$.
Conducting-sheath boundary conditions are applied to $f$ in $z$,
which partially reflect gyrocenters of one species and
fully absorb gyrocenters of the other species into the wall depending on the
sign of the sheath potential.
The potential is determined by solving the gyrokinetic Poisson equation
[Eq.~(\ref{eq:gkp})].
Evaluating this potential at the sheath entrances (the ends of
the simulation domain in $z$) gives the sheath potential, which is used to
determine which particles are reflected by the sheath. This is the gyrokinetic
analog of how fluid codes have used the vorticity to calculate the potential
and sheath effects (for example, see \citet{Ricci2009a}).
We refer to these boundary conditions as conducting-sheath boundary conditions
\cite{Shi2017,ShiPhDThesis2017} because they allow self-consistent 
currents locally in and out of the end plates.
This is in contrast to the logical-sheath
boundary-condition model,\cite{Parker1993,Shi2015,Chone2018} which assumes an
insulating sheath
with zero current density at the end plates everywhere. There is no
closed-field-line region in our present model.

We use parameters roughly approximating a singly ionized H-mode deuterium
plasma in the NSTX SOL:\cite{Zweben2015,Zweben2016}
$n_{i0}^g = 7 \times 10^{18}$~cm$^{-3}$, $T_e \sim 40$~eV, $T_i \sim 60$~eV, $B_{\mathrm{axis}} = 0.5$~T, $R_0 = 0.85$~m,
$a_0 = 0.5$~m.
Although we use parameters for an H-mode plasma, we do not attempt or claim to capture H-mode physics (e.g.
an edge transport barrier) in our simulations, since they include only the SOL
and not the pedestal.

The simulation box has dimensions $L_x = 50 \rho_{\mathrm{s}0} \approx 14.6$~cm, $L_y = 100 \rho_{\mathrm{s}0} \approx 29.1$~cm,
$L_z = L_p/\sin \theta$, where $L_p = 2.4$~m, $\rho_{\mathrm{s}0} = c_{\mathrm{s}0}/\Omega_i$,
and $\theta$ is the magnetic-field-line incidence angle.
The magnetic field is taken to be comprised primarily of a toroidal component with
a smaller vertical component (referred to as $B_v$), resulting in a helical-field-line geometry
that roughly approximates the tokamak SOL.
We present results with $\sin \theta = B_v/B_z = \left ( 0.2, 0.3, 0.6 \right)$ in
Sec.~\ref{sec:results}, which correspond to $L_z = \left(12,8,4 \right)$~m.
The connection length to the divertor plate in the real NSTX experiment is
typically quite long, over 10~m, but we consider smaller values that might
represent the shorter connection length from the midplane to the X-point
region, where the magnetic shear is very strong.
In this study, the magnetic-field-line incidence angle is not accounted for in the sheath boundary conditions
(i.e. no Chodura sheath \citep{Chodura1982}).

We use an energy-conserving discontinuous Galerkin method for the spatial
discretization of the equations,
which is a generalization of the algorithm of Liu \& Shu \cite{Liu2000} for
two-dimensional incompressible flow
in the vorticity--stream function formulation.
Time discretization is performed using an explicit third-order
strong-stability-preserving Runge--Kutta algorithm.\cite{Gottlieb2001}
The positivity of the distribution function is not automatically guaranteed
in our algorithms, and our method
to keep $f > 0$ results in the addition of a small amount of numerical 
heating $\sim$10\% of the source power to the system.
The details of the numerical algorithms, energy conservation, 
and sheath boundary conditions
are discussed in Refs.~\onlinecite{Shi2017} and \onlinecite{ShiPhDThesis2017}.

The position-space extents are $x \in [R_0 + a_0 - L_x/2, R_0 + a_0 + L_x/2]$, $y \in [-L_y/2, L_y/2]$, $z \in [-L_z/2, L_z/2]$,
and the velocity-space extents are $v_{\parallel s} \in
[-v_{\parallel s,\mathrm{max}},
v_{\parallel s, \mathrm{max}}]$,
where $v_{\parallel s,\mathrm{max}} = 4 v_{ts} = 4 \sqrt{T_s/m_s}$
and $\mu_{s,\mathrm{max}} = (3/4) m_s v_{\parallel s,\mathrm{max}}^2 / (2B_0)$,
where $B_0 = B_{\mathrm{axis}}R_0/(R_0+a_0)$.
The solution in each cell is expanded using piecewise-linear basis functions, i.e. the
span of monomials in the five phase-space variables with each variable degree $\le 1$.
This choice results in 32 degrees-of-freedom per element to represent the distribution function and Hamiltonian.
The grid resolution is $(N_x,N_y,N_z,N_{v_\parallel},N_\mu) = (18,36,10,10,5)$, and a uniform grid spacing is used.

The plasma density source has the following form:
\begin{equation}
S(x,z) = \left\{ 
  \begin{array}{ll}
  S_0 \mathrm{max}\left[ \exp\left(\frac{-(x-x_S)^2}{2 \lambda_S^2 }\right), 0.1\right], & |z| < L_z/4 \\
  0 & \mathrm{else},
  \end{array} \right. \label{eq:helical_sol_source}
\end{equation}
where $x_S = -0.05 \; \mathrm{m} + R_0 + a_0$, $\lambda_S = 5 \times 10^{-3}$~m,
and $S_0$ is chosen so that the source has total (electron plus ion) power
$P_{\mathrm{source}} = 0.27 L_z/L_{z0} $ MW, where $L_{z0} = 4$~m.
The expression for the source power comes from multiplying $P_{\mathrm{SOL}} = 5.4$~MW, the total power into the SOL,
by the fraction of the total device volume covered by the simulation box.
A floor of $0.1 S_0$ is used in the $|z| < L_z/4$ region to prevent regions of $n \ll n_0$ from developing at large $x$,
which can result in distribution-function positivity issues.
The distribution function of the sources are non-drifting Maxwellians
with a temperature profile $T_{e,i} = 74$~eV for $x < x_S + 3 \lambda_S$ and $T_{e,i} = 33$~eV for $x \ge x_S + 3 \lambda_S$.
These choices result in an integrated source particle rate of ${\approx}9.6 \times 10^{21}$~s$^{-1}$ for
the $L_z = L_{z0}$ ($B_v/B_z = 0.6$) case.

We do not yet include a closed-field-line region in our simulations, 
so we only simulate a SOL.
The $x < x_S + 3 \lambda_S$ region will be referred to as the source region in this paper,
while the $x \ge x_S + 3 \lambda_S$ region will be referred to as the SOL region.
We can think of the $x = x_S + 3 \lambda_S$ location as the LCFS.

\section{Simulation Results \label{sec:results}}
\begin{figure}
\includegraphics[width=\linewidth]{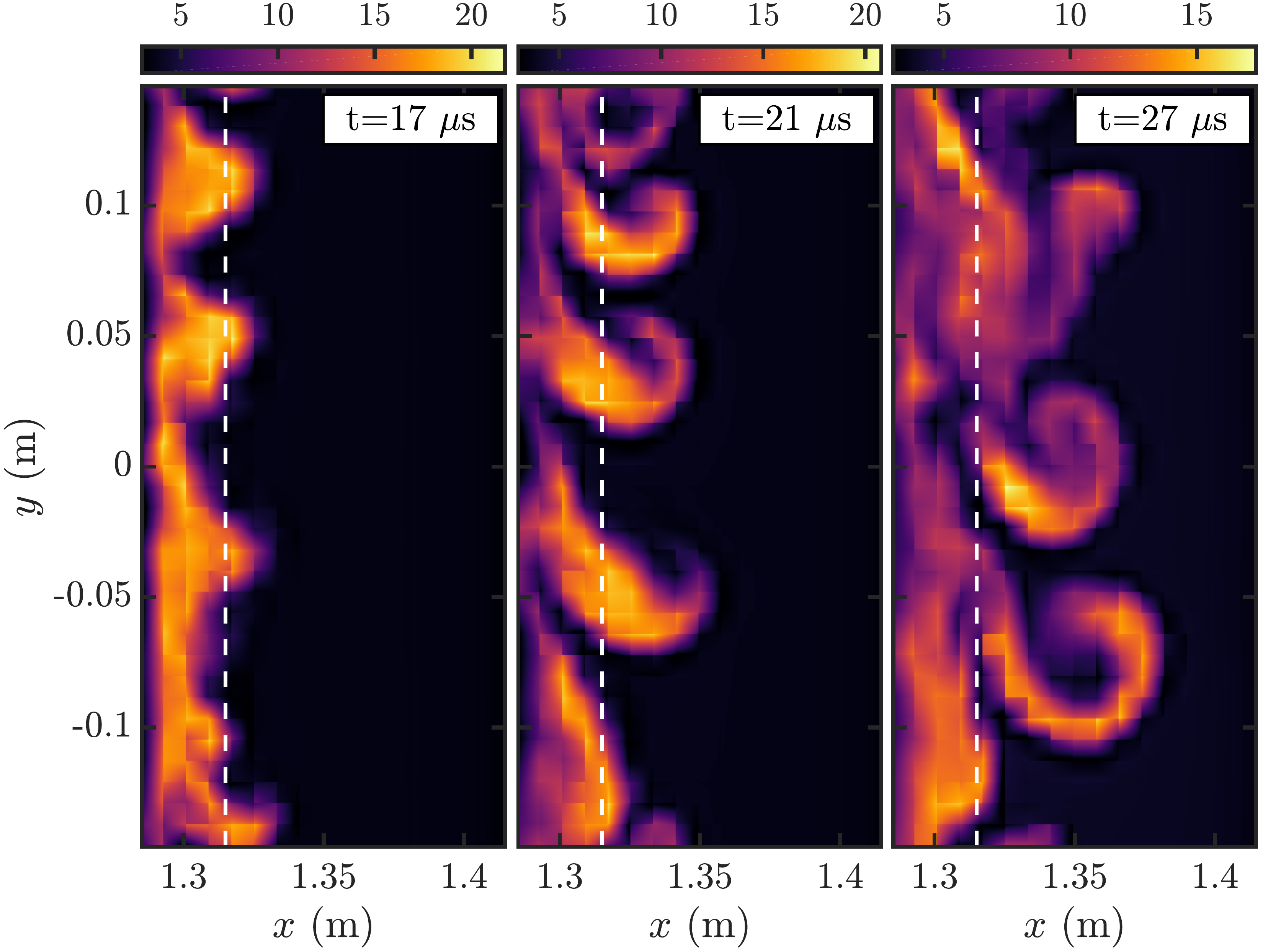}
\caption{\label{fig:blob_formation} Snapshots of the electron density (in 10$^{18}$ m$^{-3}$)
  at various times ($t=17$~$\mu$s, 21~$\mu$s, and 27~$\mu$s) near the beginning of a simulation
  in the perpendicular $x$--$y$ plane at $z=0$~m.
  This simulation has $B_v/B_z = 0.3$.
  The dashed line indicates the boundary between the source and SOL regions.
  Note that each plot uses a different color scale to better
  show the features.}
\end{figure}

\begin{figure}
\includegraphics[width=\linewidth]{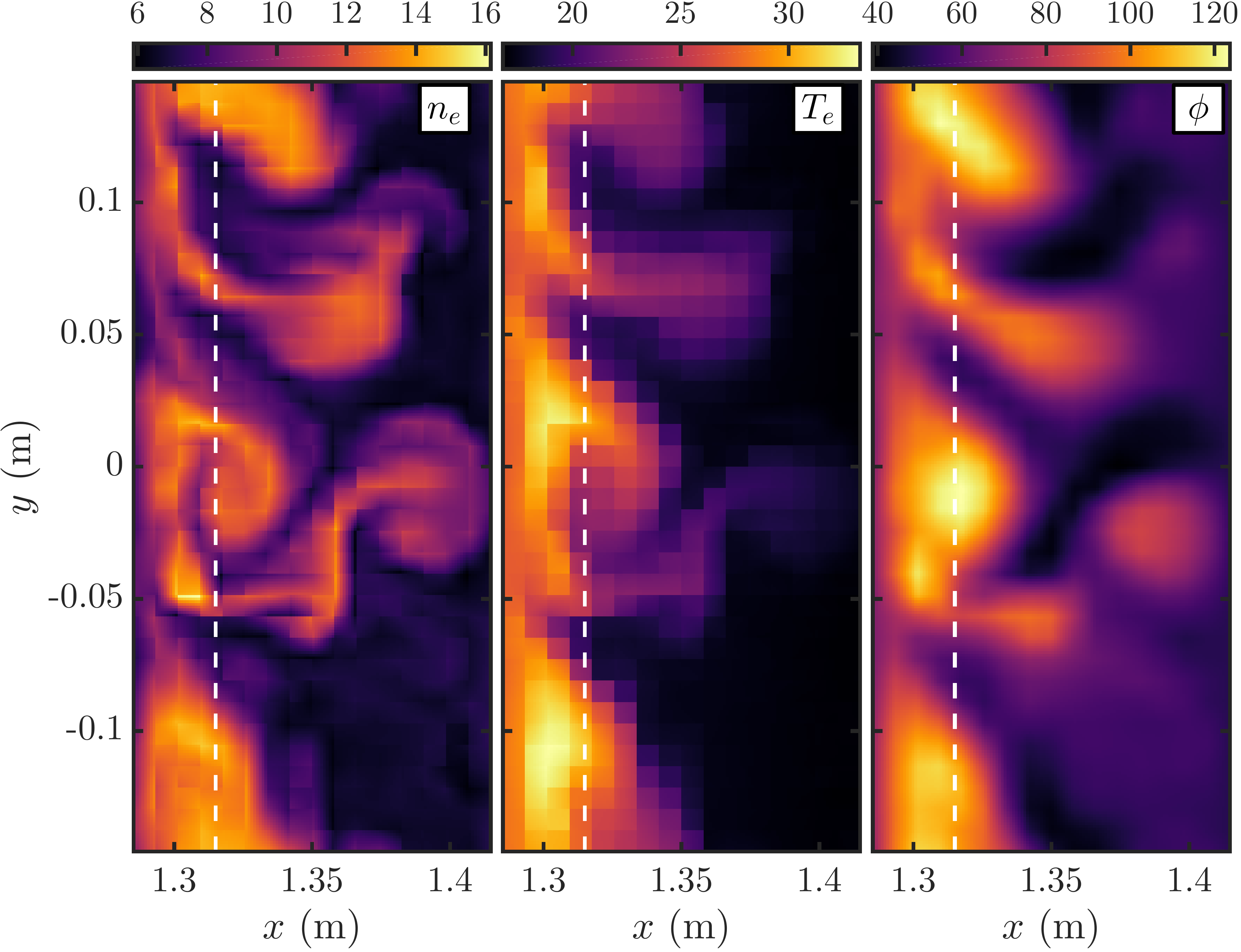}
\caption{\label{fig:test145} Snapshots of the electron density (in 10$^{18}$~m$^{-3}$), electron temperature
  (in eV), and electrostatic potential (in V) in the plane perpendicular to the magnetic field
  at $z=0$~m. This plot is made at $t=625$~$\mu$s, which is after several ion
  transit times ($\tau_i \sim 50$~$\mu$s). This simulation has $B_v/B_z = 0.3$.
  The dashed line indicates the boundary between the source and SOL regions.
  A mushroom structure in the electron density is observed at large $x$.}
\end{figure}

Starting from an initial condition estimated by the steady-state solution of 
one-dimensional fluid equations
(see Appendix~\ref{sec:initial-helical-sol}),
the sources steepen the plasma profiles, quickly triggering curvature-driven modes that grow on a timescale
comparable to $\gamma \sim c_s/\sqrt{R \lambda_p}$.
We emphasize that our system does not contain ballooning modes since there are no `good-curvature' regions.
As shown in Fig.~\ref{fig:blob_formation}, radially elongated structures extending far from the source region
are generated and subsequently broken up by sheared flows in the $y$ direction in the source region,
leaving radially propagating blobs in the SOL region.
Using the time-averaged profiles from the same $B_v/B_z = 0.3$ ($L_z = 8$~m) simulation, we
estimate $\gamma \sim 1.9 \times 10^5$~s$^{-1}$ using
 $\lambda_p \approx 2.9$~cm, $T_e \approx 30$~eV, and $R = x_S = 1.3$~m.
On a time scale long compared to $\gamma^{-1}$ and $\tau_i = (L_z/2)/v_{ti}
\sim 50$~$\mu$s, the conducting-sheath boundary conditions
maintain a quasi-steady state in which the particle losses to the end plates are balanced by the plasma sources.
Snapshots of the electron density, electron temperature, and electrostatic potential from the quasi-steady state ($t=625$ $\mu$s)
for the $B_v/B_z = 0.3$ case are shown in Fig.~\ref{fig:test145}.

\begin{figure}
\includegraphics[width=\linewidth]{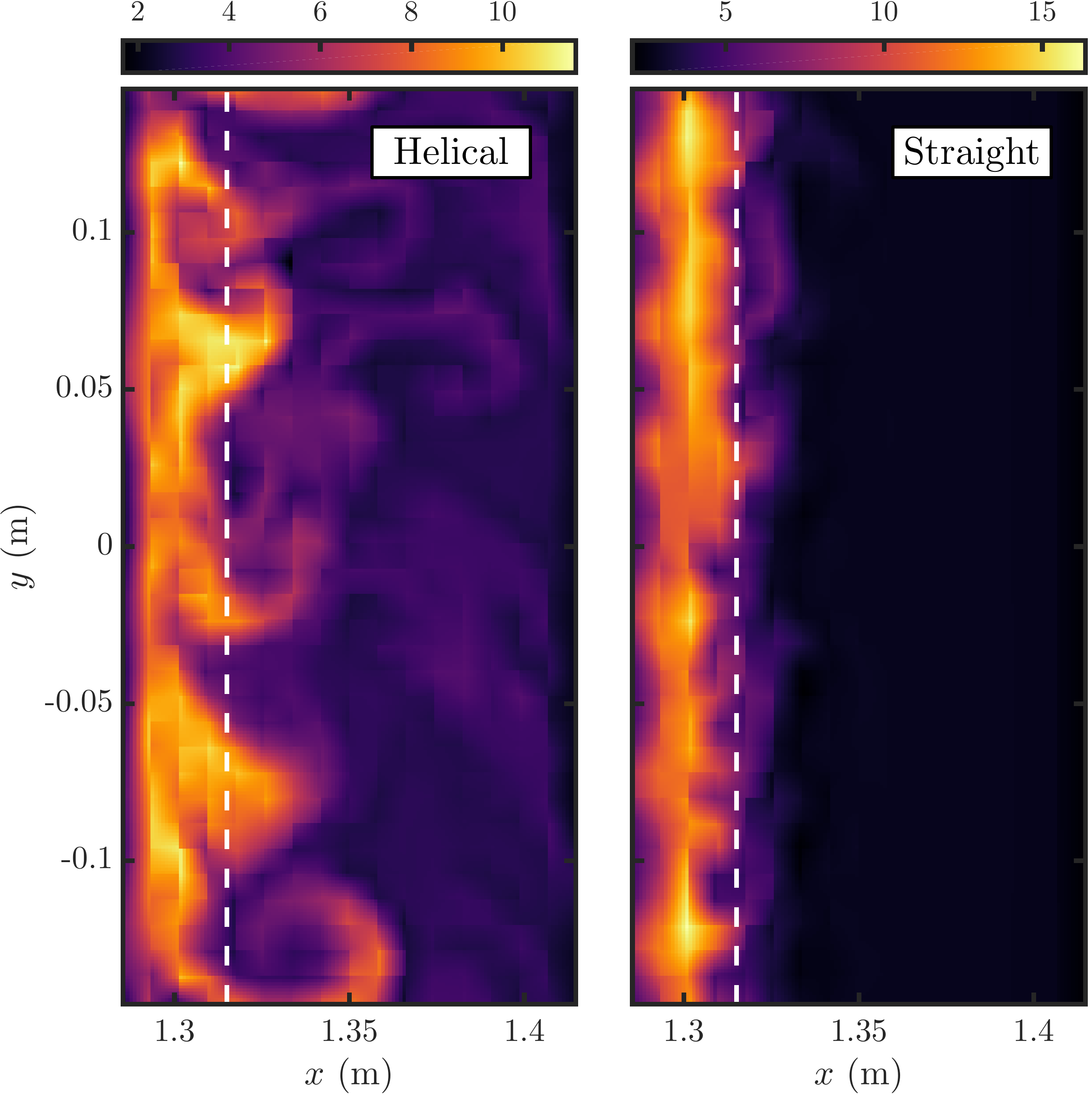}
  \caption{Comparison of an electron density snapshot (in $10^{18}$~m$^{-3}$) between  a simulation
  in a helical-magnetic-field-line geometry and a simulation in a straight-magnetic-field-line
  geometry with $B = B_0$.
  The formation of blobs in the helical-SOL simulation results in the transport of density
  to large $x$ and a broad density profile.
  Coherent structures of elevated plasma density do not appear to convect to
  large $x$ in the straight-magnetic-field-line case, and so density is 
    mostly localized to the source region.
  The plots are made in the perpendicular $x$--$y$ plane at $z = 0$~m and $t = 681$~$\mu$s.
  The dashed line indicates the boundary between the source and SOL regions.
  Note that each plot uses a different color scale to better show the features.}
  \label{fig:curvature_comparison}
\end{figure}

For the steepest magnetic-field-line-pitch case ($B_v/B_z = 0.6$),
we performed a second simulation with magnetic-curvature effects removed,
keeping all other parameters unchanged.
The resulting magnetic geometry consists only of straight magnetic field lines,
so coherent structures of elevated
plasma density cannot become polarized by curvature forces.
As shown in the electron-density snapshot comparison in Fig.~\ref{fig:curvature_comparison},
the presence of magnetic curvature appears to have
an important role in the turbulent dynamics of the system.
When magnetic-curvature effects are removed, the radial propagation of
coherent structures into the SOL region appears to be significantly weakened or absent,
and most of the density is localized to the source region.

\begin{figure*}
\includegraphics[width=\linewidth]{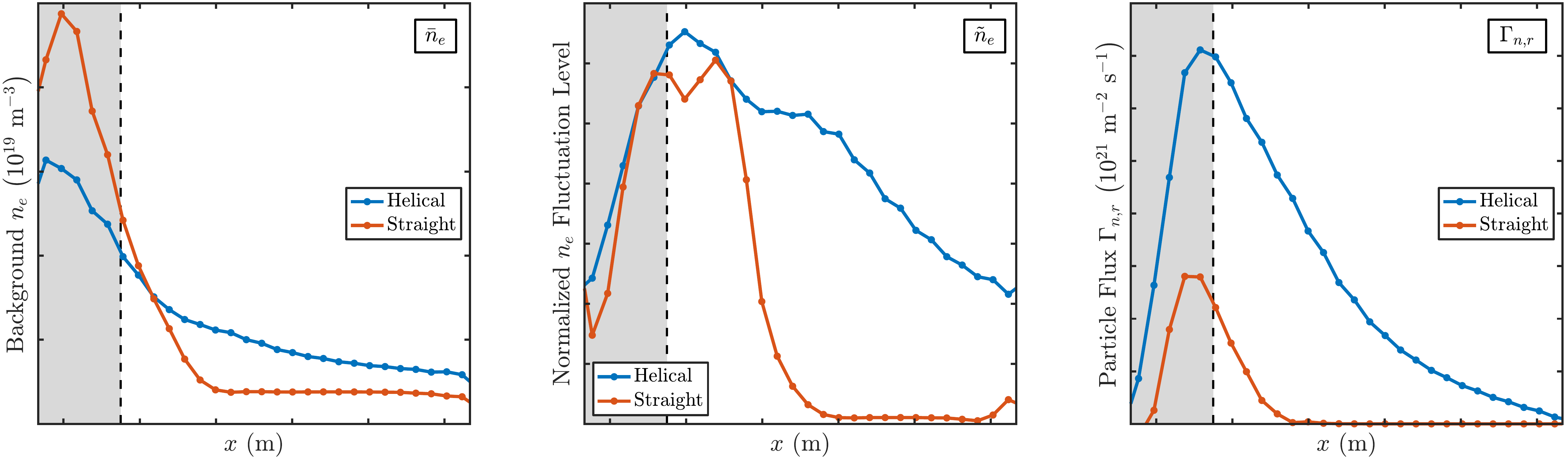}
  \caption{Radial profiles of the background electron densities (in $10^{19}$~m$^{-3}$),
  normalized electron-density fluctuation levels, and radial $E \times B$ particle fluxes $\Gamma_{n,r}$
  (in $10^{21}$~m$^{-2}$~s$^{-1}$) for
  a helical SOL simulation and a straight-field-line simulation with $B = B_\mathrm{axis}$.
  These plots are computed using data near the midplane in the region -0.5~m$<z<0.5$~m and 
  sampled at 0.25~$\mu$s intervals over a ${\sim}400$~$\mu$s period.
  The shaded area indicates the region in which the source is concentrated.
  The background density profile in the straight-field-line simulation does not decay to 0
  at large $x$ due to the presence of a constant low-amplitude source in that region to help mitigate
  positivity issues with the distribution function.}
  \label{fig:curvature_comparisons_1d}
\end{figure*}

Figure~\ref{fig:curvature_comparisons_1d} compares radial profiles of the background electron densities,
normalized electron-density fluctuation levels, and radial $E \times B$ particle fluxes $\Gamma_{n,r}$
between these two simulations.
The radial particle flux due to electrostatic turbulence is calculated
as $\Gamma_{n,r} = \langle \tilde{n}_e \tilde{v}_r \rangle$,\citep{Zweben2007}
where $v_r = E_y/B$ and the brackets $\langle \dots \rangle$ indicate an 
average over a period that is long compared to the fluctuation time scale
and an average over $y$ and the central region in $z$, -0.5~m$<z<0.5$~m.
The fluctuation of a time-varying quantity $A$ is denoted as $\tilde{A}$, 
which is related to the total $A$ as $\tilde{A} = A - \langle A \rangle_t$.
Here, the brackets $\langle \dots \rangle_t$ indicate an average in time.
Notable differences between these two simulations are found in all three quantities plotted.
Compared to the helical-SOL simulation, the straight-field-line simulation has a background density
profile that decays more rapidly, fluctuation levels that quickly drop to
${\approx}0\%$ outside
$x \approx 1.35$~m, and a ${\approx}2.5$ times smaller $\Gamma_{n,r}$ that also
drops to approximately zero outside $x \approx 1.34$~m.

We have also performed a scan of the mass ratio $m_i/m_e$ from the actual ratio of 3698 down to 100
(by increasing the electron mass), and we observed no significant quantitative or
qualitative changes in the turbulence.
The mass ratio might play an important role in a different parameter regime, however.

Effects connected to $B_v \sim B_p \sim I_{\mathrm{plasma}}$ are explored by changing
the magnetic-field-line incidence angle, since $\sin\theta = B_v/B_z$.
We have performed simulations at three values of magnetic-field-line pitches
$B_v/B_z = (0.2,0.3,0.6)$, which correspond to $L_z = (12,8,4)$~m and
$\theta = (20.14^\circ,30.47^\circ,64.4^\circ)$.
We scale the source appropriately in each simulation to keep the volumetric source rate the same.
In all these simulations, the source is localized to the $z\in \left[-L_z/4, L_z/4\right]$ region
to model a source with a fixed poloidal extent.
As $\theta$ is decreased, the plasma profiles are observed to become less peaked, implying
that turbulence transport in the $x$-direction increased with decreasing $\theta$.

\begin{figure}
  \centering
  \includegraphics[width=\linewidth]{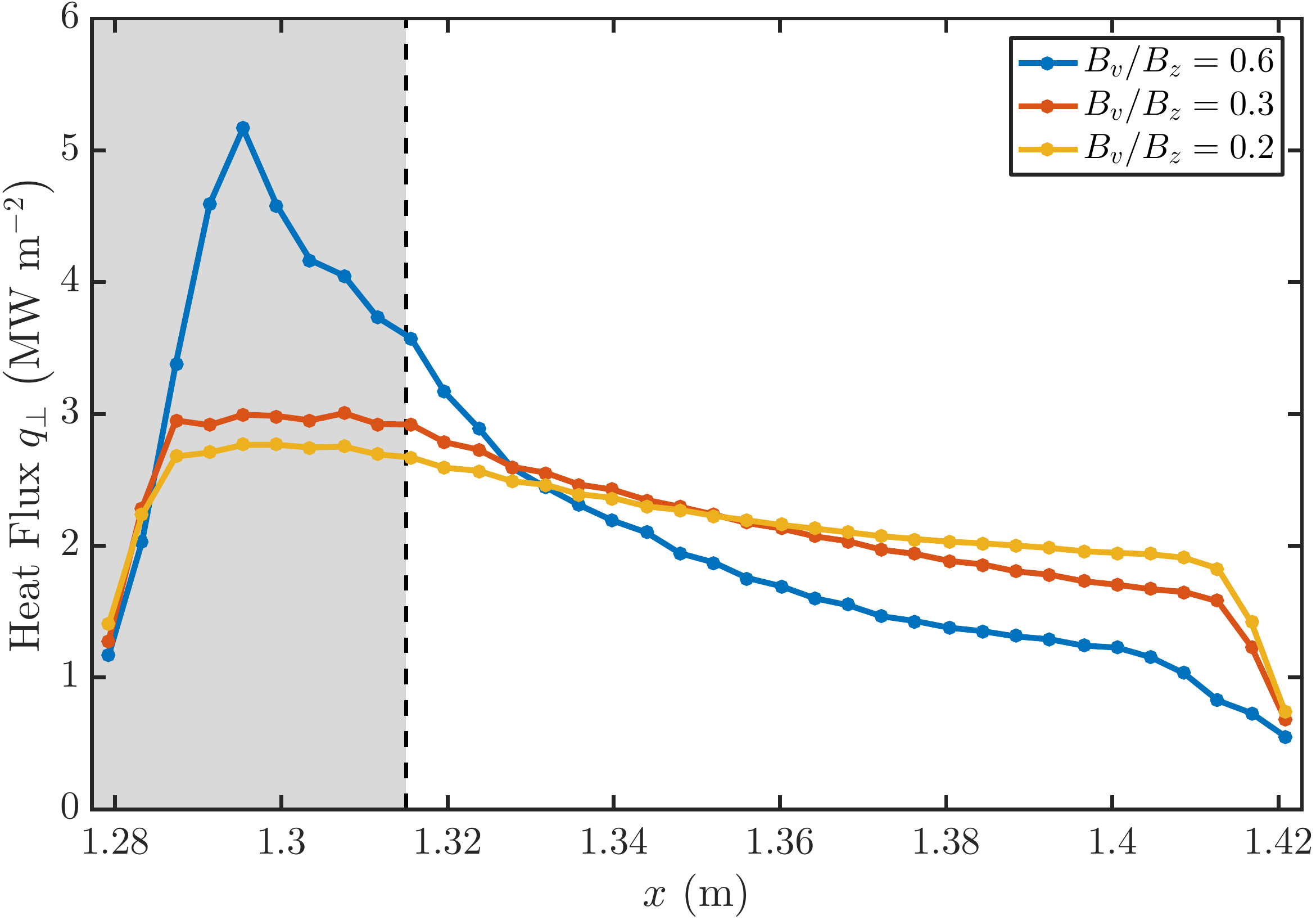}
  \caption{Comparison of the steady-state parallel heat flux normal to the divertor plate for three cases with
  different magnetic-field-line pitches. The shaded area indicates
  the region in which the source is concentrated. The heat-flux profile is observed
  to broaden as $B_v/B_z$ is decreased.
  Since a large amount of plasma gets near the outer radial boundary, where 
    further radial transport is suppressed by the $\phi = 0$ constant
    ideal-conducting-wall condition, the profiles in the shallower-pitch 
    cases may exhibit more of a uniform exponential fall off by
    increasing the box size in the radial direction.}
  \label{fig:helical-heat-flux} 
\end{figure}

We calculate the steady-state parallel heat flux $q = \sum_s \int \mathrm{d}^3 v \, H_s v_\parallel f_s$
at the sheath entrance and average $q$ in the $y$-direction to obtain a radial profile of the steady-state parallel heat flux
for each case.
To compare the heat fluxes on an equal footing, we plot the component of the parallel heat flux normal to the divertor plate $q_\perp = q \sin \theta$
in Fig.~\ref{fig:helical-heat-flux}.
Compared to the $B_v/B_z = 0.6$ case, the heat-flux profiles for the cases with a shallower pitch
are much broader.
This behavior is consistent with the observation in tokamaks that the SOL heat-flux width is inversely
proportional to the poloidal magnetic field (analagous to $B_v$ in this model) and the plasma
current,\citep{Eich2013,Makowski2012} although the physical reasons behind the scaling in our model
and in a tokamak SOL may be quite different.
We note that a significant amount of plasma in the smallest $\theta$ case
gets near the outer radial wall, where further radial transport is suppressed,
since the outer boundary is taken to be an ideal conducting plate with
constant $\phi$, so the $\bf{E} \times \bf{B}$ velocity into the side walls,
$\propto \partial \phi / \partial y$, vanishes. 
Simulations with a larger domain extent in the $x$ coordinate (and/or
finite-Larmor-radius effects in the collision operator to include classical
transport to the side wall) might exhibit more of an exponential fall off over
a wider radial range, further reducing the density in the right-hand side of
the simulation.

\begin{figure}
\includegraphics[width=\linewidth]{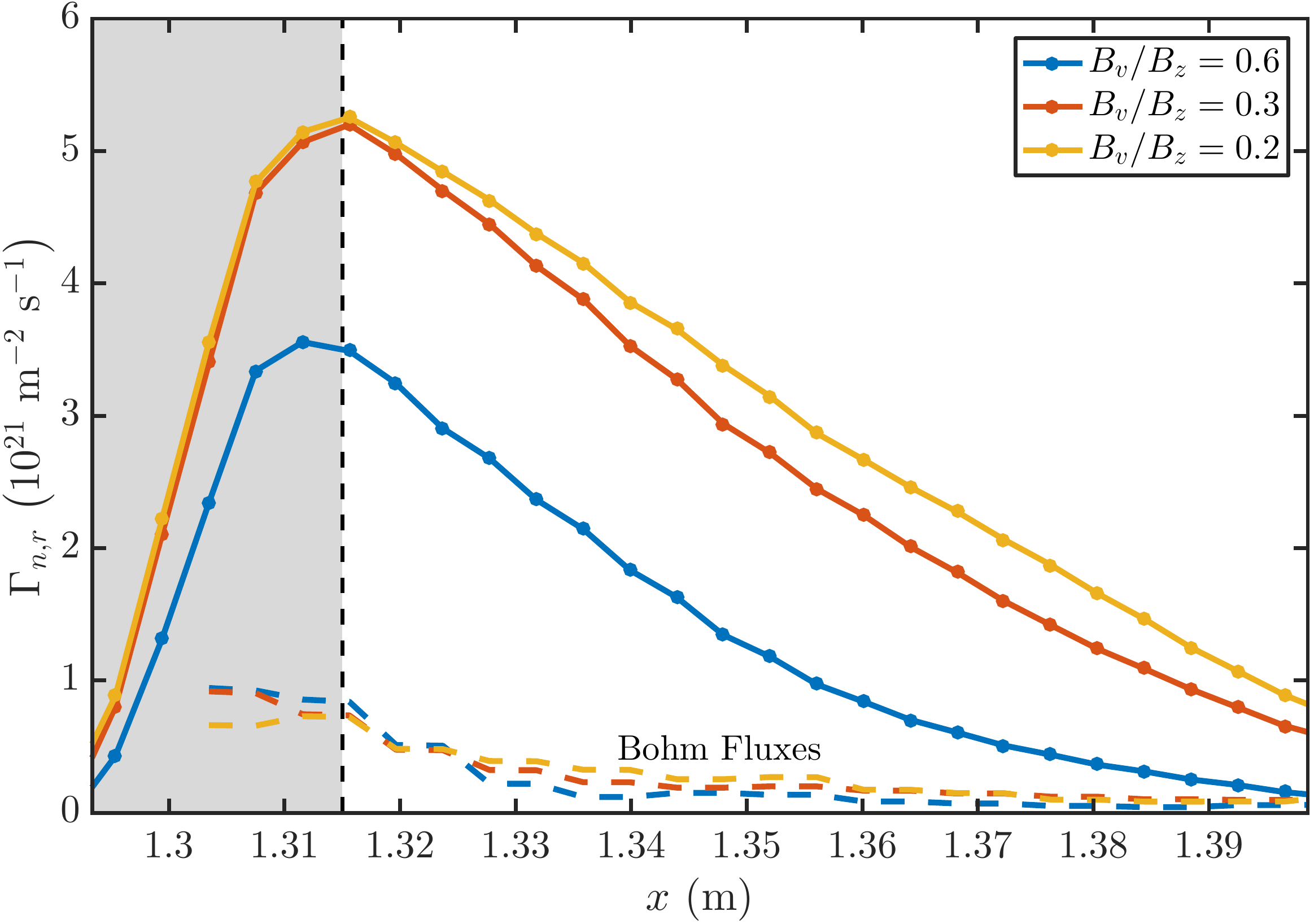}
\caption{\label{fig:helical-particle-flux} Comparison of the radial $E \times B$ particle flux evaluated
  near the midplane for three cases with different magnetic-field-line pitches.
  The shaded area indicates the region in which the source is concentrated.
  The dashed lines are Bohm-flux estimates for comparison.}
\end{figure}

The broad heat-flux profiles in Fig.~\ref{fig:helical-heat-flux} can be connected to the increased
outward radial turbulent transport as $B_v/B_z$ becomes shallower.
We compute the steady state radial particle flux $\Gamma_{n,r}$ near the midplane
in the region -0.5~m$<z<0.5$~m for each value of $B_v/B_z$ and plot the $y$-averaged fluxes in
Fig.~\ref{fig:helical-particle-flux} (solid lines).
Since the simulation box occupies a larger fraction the device volume as $B_v/B_z$ is decreased, but
the source occupies the same fraction of the simulation box and has a fixed volumetric source rate,
the background density levels increase as $B_v/B_z$ decreases.
Another way to say this is that with a fixed volumetric source density (fixed
in particles per cubic meter per second), the mean density is expected to
increase as the parallel connection length $L_z/2$ increases and the parallel
loss rate ${\sim} 1/\tau_i = 2 v_{ti}/L_z$ decreases.
Therefore, the magnitude of the $\Gamma_{n,r}$ profiles in Fig.~\ref{fig:helical-particle-flux}
should not be taken alone as a measure of turbulence levels.

The $\Gamma_{n,r}$ profiles can be compared with the radial particle fluxes that result from
assuming Bohm diffusion, i.e. $\Gamma_B = D_B \partial_x n_e$, where the diffusion coefficient
$D_B = (1/16) k_B T_e/(eB)$.  In the $x > 1.36$~m region,
$\Gamma_{n,r}/\Gamma_B \approx 16$ for the $B_v/B_z = 0.2$ case,
while $\Gamma_{n,r}/\Gamma_B \approx 8$ for the $B_v/B_z = 0.6$ case.
One might expect the maximum level of turbulent transport to be comparable to the levels set by
$D_B$, but it is important to remember that $D_B$ is a \textit{diffusive} transport estimate.
The \textit{convective} transport of blobs in these simulations appears to be responsible for
the much-higher turbulent fluxes.
Experimental data from tokamaks also suggest that the higher-than-Bohm particle transport
in the SOL is due to the non-diffusive transport of blobs.\citep{Krasheninnikov2008,Zweben2007}

\begin{figure*}
\includegraphics[width=\linewidth]{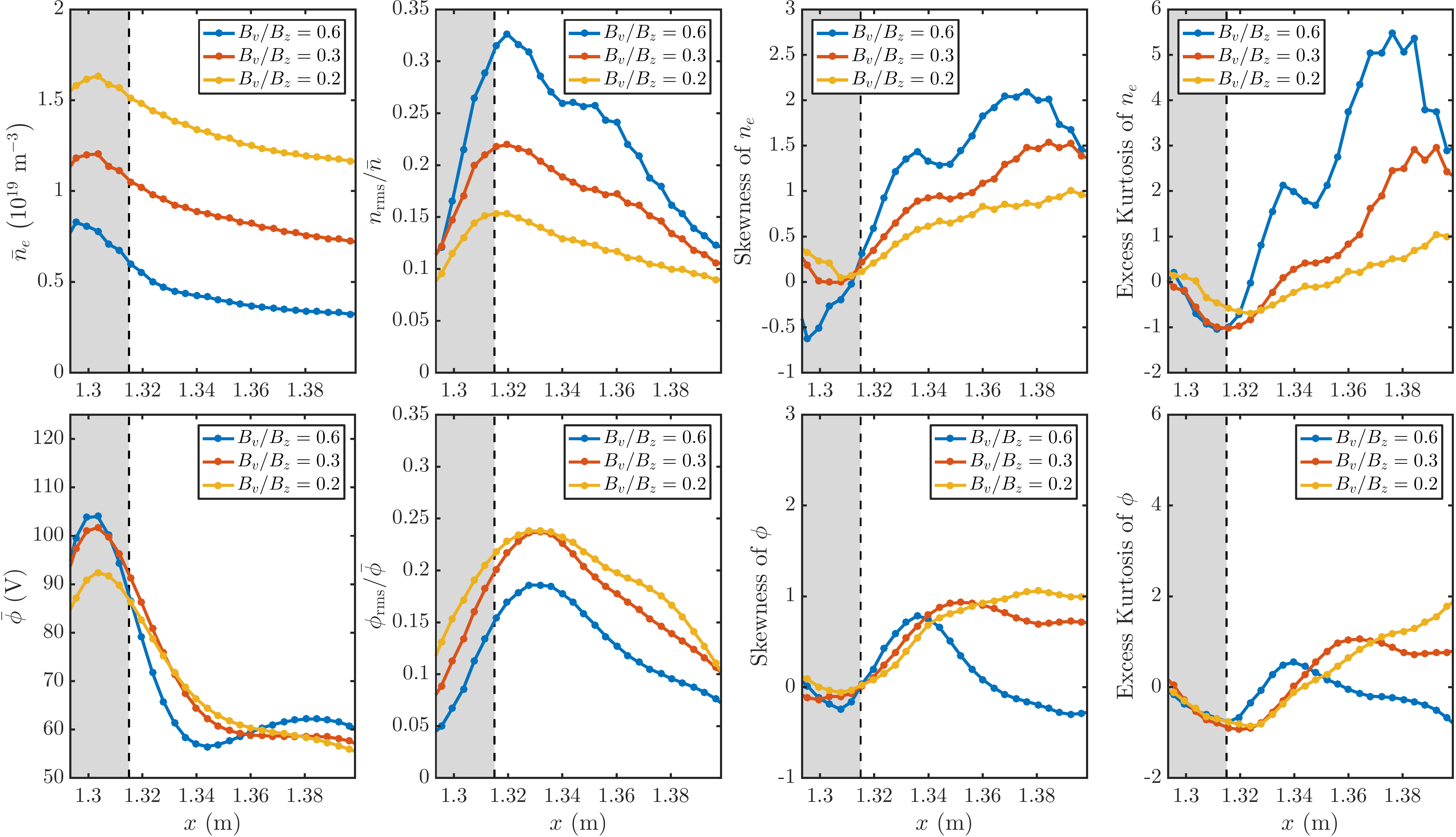}
\caption{\label{fig:n_and_phi_statistics} Comparison of the electron-density fluctuation statistics (top row)
  and electrostatic-potential fluctuation statistics (bottom row) computed near the $z=0$~m plane for three cases with
  different magnetic-field-line pitches. The potential fluctuations are notably
  less intermittent than the density fluctuations. The shaded area indicates
  the region in which the source is concentrated.}
\end{figure*}

Density fluctuation statistics are often of interest in the SOL to characterize the turbulence.
Considering again a time-varying quantity $A$, we define the skewness of $A$ as $E[\tilde{A}^3]/\sigma^3$
and the excess kurtosis of $A$ as $E[\tilde{A}^4]/\sigma^4 - 3$, where $\sigma$ is the standard
deviation of $A$ and $E[\dots]$ denotes the expected value.
Figure~\ref{fig:n_and_phi_statistics} shows the radial profiles of the normalized fluctuation level,
skewness, and excess kurtosis for electron-density fluctuations and electrostatic-potential fluctuations computed
near the $z=0$~m plane.
The density and potential fluctuations are normalized to their local background values.
The positive skewness and excess kurtosis values are signatures of intermittency, which indicates
an enhancement of large-amplitude positive-density-fluctuation events and is connected to the
transport of blobs.\citep{Zweben2007,Krommes2008}

A somewhat counter-intuitive result is the reduction of density fluctuation levels as $B_v/B_z$ is decreased,
given that Figs.~\ref{fig:helical-heat-flux} and \ref{fig:helical-particle-flux} indicate that turbulent
spreading is increased as $B_v/B_z$ is decreased.
The skewness and excess kurtosis plots in Fig.~\ref{fig:n_and_phi_statistics} indicate that
the density fluctuations become closer to a normal distribution as $B_v/B_z$ is decreased.
These trends in the density fluctuation statistics can be understood by noting that
the background density profile becomes less peaked and more uniform in the $x$-direction as $B_v/B_z$ is decreased,
so a blob that is formed in the source region propagating in the SOL has a density
that is closer to the background level, which results in lower relative fluctuation,
skewness, and excess kurtosis values when compared to the large $B_v/B_z$ case.
Additionally, the density flux is constrained by the use of a fixed volumetric source rate,
so as the background density increases with decreasing $B_v/B_z$,
the relative density fluctuation levels tend to decrease.
We also observe that the potential fluctuations are much less intermittent than the density fluctuations
at the same $B_v/B_z$.
This observation could be a real, physical effect, but we note that the fact that the temperature at large $x$
runs into the grid resolution (the lowest temperature that can be represented on the velocity grid)
could be influencing the potential fluctuation statistics in this region.
Unlike the density fluctuations, the normalized potential fluctuation levels tend to increase with decreasing
$B_v/B_z$.

\begin{figure*}
\includegraphics[width=\linewidth]{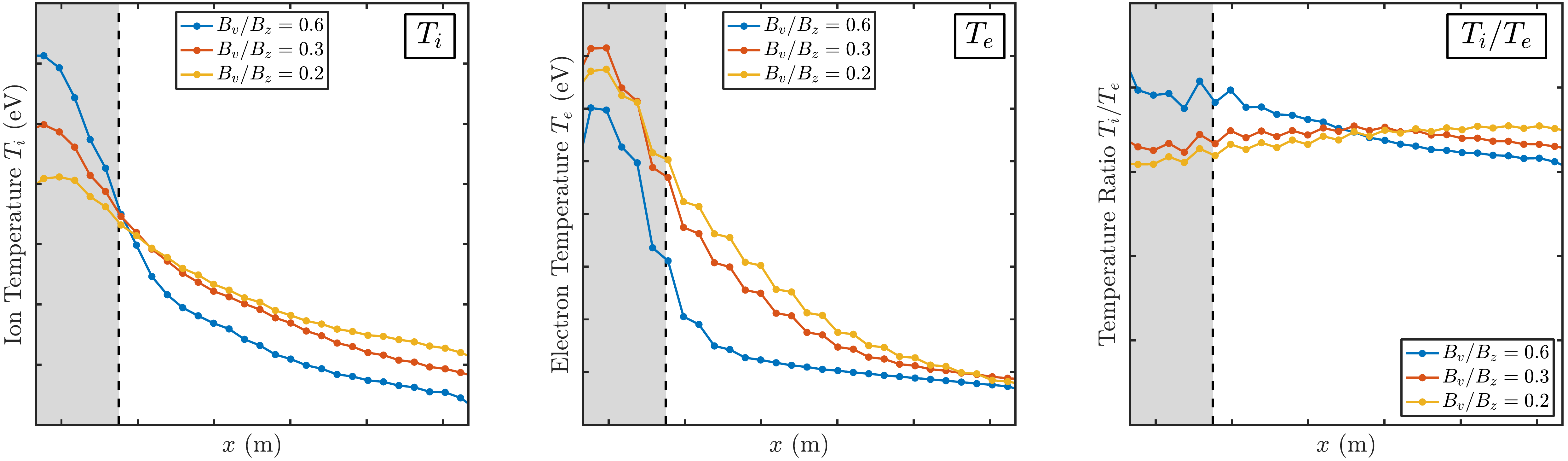}
  \caption{Radial profiles of the steady-state ion and electron temperatures near the midplane
  and ion-to-electron temperature ratios for cases with different magnetic-field-line pitches.
  Although both electrons and ions are sourced at the same temperature, the sheath allows
  high-energy electrons to be rapidly lost from the system, resulting in lower
  electron temperatures in the SOL if collisions are not rapid enough to equilibrate 
  the two species.\citep{Stangeby1990,Kocan2011}}
  \label{fig:helical-sol-temperatures} 
\end{figure*}

Figure~\ref{fig:helical-sol-temperatures} shows radial profiles of the
steady-state ion and electron temperatures and ion-to-electron
temperature ratios near the midplane for
different $B_v/B_z$.
For all three simulations, $T_i/T_e$ falls in the range 1.5--2, which is within the range of
1--10 that is observed a few centimeters outside the LCFS in tokamaks.\citep{Kocan2011}
Similar to the heat-flux profiles shown in Fig.~\ref{fig:helical-heat-flux},
the profiles are steepest for the case with $B_v/B_z=0.6$ and decay
more gradually in the lower $B_v/B_z$ cases.
SOL measurements typically show that the ratio $T_i/T_e$ 
increases with radius.\citep{Kocan2011}
We see this trend in Fig.~\ref{fig:helical-sol-temperatures} for $B_v/B_z = 0.3$ and 0.2,
but not for $B_v/B_z = 0.6$.
This reversed trend for $B_v/B_z = 0.6$ is likely connected to the relatively flat
$T_e$ at large $x$.
In the $B_v/B_z = 0.6$ case, the low-amplitude source of ${\sim}33$~eV electrons at large
$x$ [see the form of the 
plasma source, Eq.~(\ref{eq:helical_sol_source})] could be setting $T_e$ in this region.

The flat $T_e$ at large $x$ could also be an artifact from the electrons
running into a floor in the temperature at large $x$.
However, we note that the minimum electron temperature allowed on our
present grid is
$T_{e,\mathrm{min}} = (2/3) T_{\perp e,\mathrm{min}} + (1/3) T_{\parallel e,
\mathrm{min}} = 11$~eV based on $T_{\perp e,\mathrm{min}}=16$~eV and
$T_{\parallel e,\mathrm{min}} = 1.1$~eV, which is somewhat lower than the
$T_e$ seen in this region.
We can test this in future work by running higher-resolution runs, including
a variable $\mu$ grid to better resolve low energies or by using exponential
reconstructions, which is currently being added to the code.

\begin{figure}
\includegraphics[width=\linewidth]{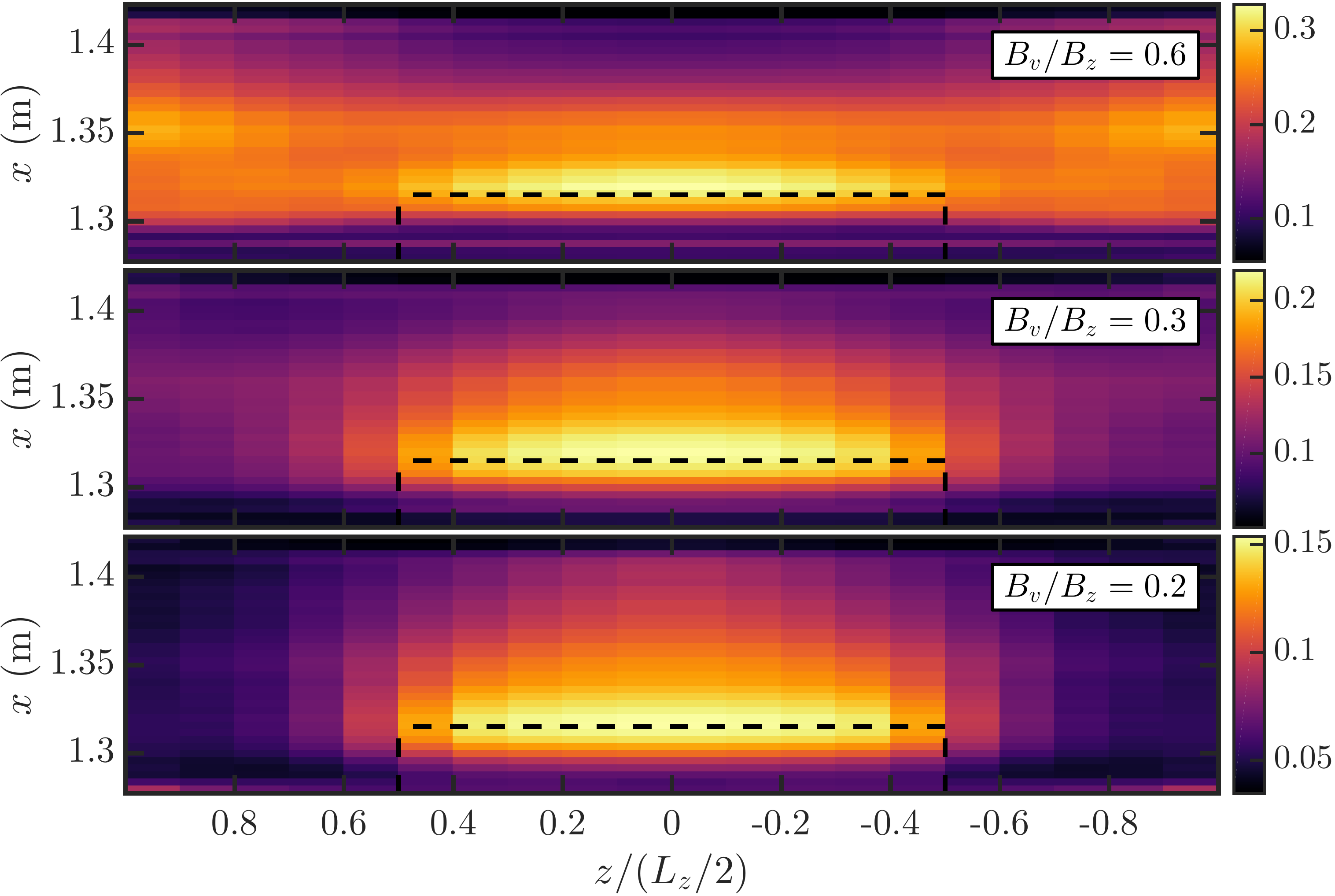}
\caption{\label{fig:parallel_rms_density} Comparison of the parallel structure of the normalized r.m.s. electron-density fluctuation amplitude
  for three cases with different magnetic-field-line pitches.
  While the density fluctuations are primarily $k_\parallel =0$ in the $B_v/B_z = 0.6$ case,
  more parallel structure is observed in the lower $B_v/B_z$
  cases. The region in which the source is concentrated is indicated by the dashed black lines.}
\end{figure}

The normalized root-mean-square (r.m.s.) electron-density fluctuation level in the $x$--$z$ plane is shown
in Fig.~\ref{fig:parallel_rms_density}.
For all three values of $B_v/B_z$, the density fluctuation levels are the largest in the source region
$|z| < L_z/4$.
The normalized density fluctuation levels in the $B_v/B_z = 0.6$ case are fairly uniform along the
field lines, while they tend to fall off by about a factor of 2--3 towards the sheaths in the
smaller $B_v/B_z$ cases.
This effect could be a result of the stronger influence of the sheath on
the potential as the distance from the source to the sheath is decreased.
The instantaneous snapshots of $\tilde{n}_e$ (not shown) indicate a strong $k_\parallel =0$ component for the
largest $B_v/B_z$ cases, while more parallel structure is apparent in the smaller $B_v/B_z$ cases.

\begin{figure*}
  \includegraphics[width=\linewidth]{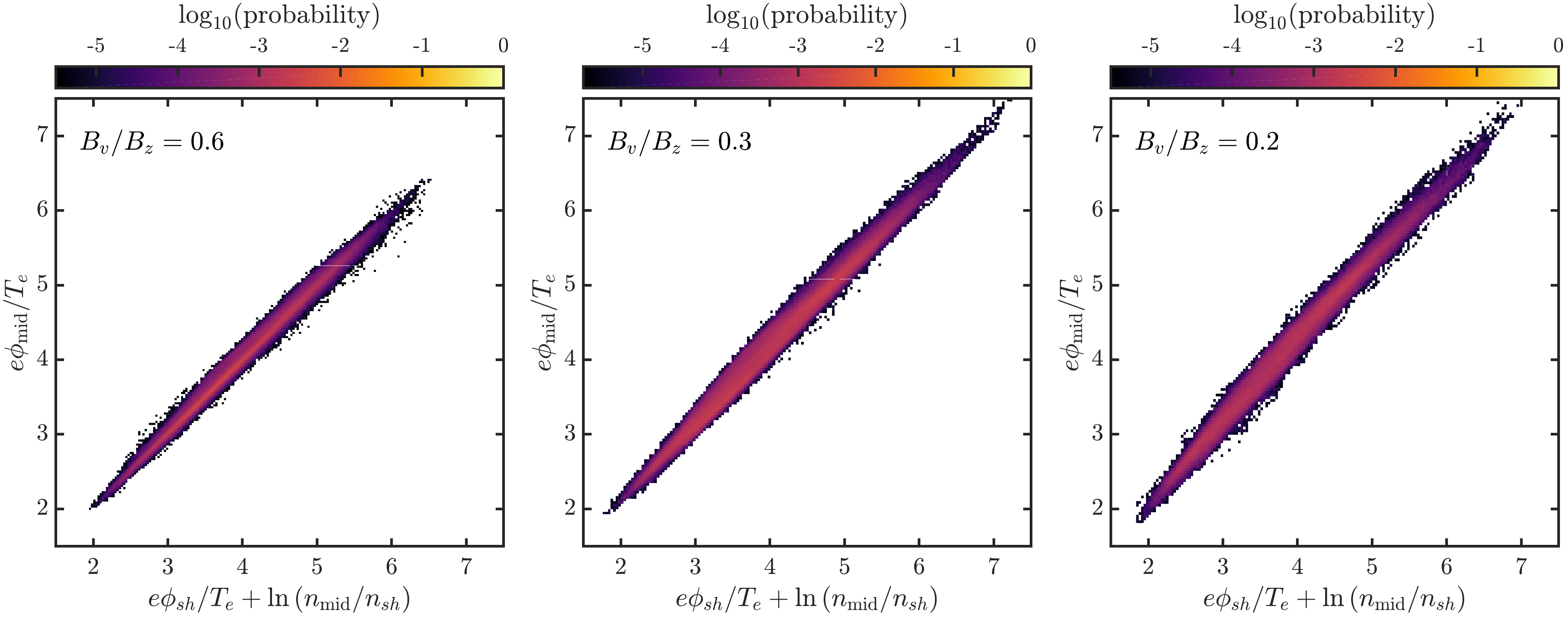}
  \caption{Comparison of the cross-coherence between the midplane potential $e\phi_{\mathrm{mid}}/T_e$ and
    $e\phi_{sh}/T_e + \ln \left(n_{\mathrm{mid}}/n_{sh} \right)$ 
    [see Eqs.~(\ref{eq:force_balance})--(\ref{eq:cross_coherence})] 
    for three cases with different magnetic-field-line pitches $\theta$.
    Here, $\phi_{sh}$ is the sheath
  potential, $n_\mathrm{mid}$ is the midplane electron density, and $n_{sh}$ is the sheath
  electron density.
  These plots are created by binning ordered pairs of the two quantities sampled
  every 0.25 $\mu$s over a ${\sim} 1$ ms time interval at solution nodes falling in the region
  $1.318 \text{ m} \le x \le 1.326 \text{ m}$.
  In all three cases, the two quantities are highly correlated, which
  indicates that the electrons are strongly adiabatic (near parallel force balance).}
  \label{fig:helical_cross_coherence}
\end{figure*}

The fluctuation statistics can also give information about the strength of the
electron adiabatic response for each simulation. By assuming that the electrons are isothermal
along field lines, parallel force balance satisfies
\begin{eqnarray}
    -n_e e E_\parallel - \nabla_\parallel P_e &=& 0 \label{eq:force_balance}\\
  -n_e e E_\parallel &=& T_e \nabla_\parallel n_e \\
  \frac{e \nabla_\parallel \phi}{T_e} &=& \nabla_\parallel \ln n_e \\
  \frac{e \phi_\mathrm{mid}}{T_e} &=& \frac{e \phi_{sh}}{T_e} + \ln \left(
    \frac{n_{\mathrm{mid}}}{n_{sh}} \right)
  \label{eq:cross_coherence},
\end{eqnarray}
where $\phi_{sh}$ and $n_{sh}$ are the electrostatic potential and electron density evaluated
at the sheath entrances and $\phi_{\mathrm{mid}}$ and $n_{\mathrm{mid}}$ are the same quantities, but evaluated
at the midplane ($z=0$~m).
To compute the cross-coherence diagnostic,\citep{Scott2005,Ribeiro2005,Mosetto2013} ordered pairs
$\boldsymbol{(}e\phi_{\mathrm{mid}}/T_e,e\phi_{sh}/T_e + \ln \left(n_{\mathrm{mid}}/n_{sh} \right) \boldsymbol{)}$
falling in the region $1.318 \text{ m} \le x \le 1.326 \text{ m}$ (approximately
where the maximum density and potential fluctuations are) are sampled
at $1$~$\mu$s intervals over a ${\sim}1$~ms period for each simulation.
Figure~\ref{fig:helical_cross_coherence} shows the resulting plots (normalized bivariate histograms),
which all indicate a strong correlation between the two sides of
Eq.~(\ref{eq:cross_coherence}),
and so the electrons are strongly adiabatic, meaning that the electron
distribution function along a field line closely follows a Boltzmann
distribution.\citep{StoltzfusDueck2009}
This finding indicates that it might be possible to obtain similar results
using a two-dimensional turbulence model (with reduced parallel dynamics and
sheath-model boundary conditions) for the parameters considered here.
To quantify the degree of non-adiabaticity, we define the parameter
\begin{equation}
    \epsilon_\mathrm{na}^2 = \frac{E \left( \left(\bar{\phi}_\mathrm{mid}
    - \bar{\phi}_\mathrm{ad} \right)^2
    \right)}{E \left( \left[ \bar{\phi}_\mathrm{mid} -
    E\left(\bar{\phi}_\mathrm{mid} \right)
    \right]^2 \right)},
\end{equation}
where $\bar{\phi}_\mathrm{mid}$ is the left-hand side of
Eq.~(\ref{eq:cross_coherence}) and $\bar{\phi}_\mathrm{ad}$ is the
right-hand side of Eq.~(\ref{eq:cross_coherence}).
The $\epsilon_\mathrm{na}$ parameter measures of the fraction of
fluctuations in $\phi$ that are due to non-adiabatic effects.
We find that $\epsilon_\mathrm{na}$ is 0.094 for $B_v/B_z = 0.6$,
0.226 for $B_v/B_z = 0.3$, and 0.310 for $B_v/B_z = 0.2$, which is consistent
with our expectation that the electrons become less adiabatic as $B_v/B_z$ is
decreased.

\begin{figure*}
  \centering
  \includegraphics[width=\linewidth]{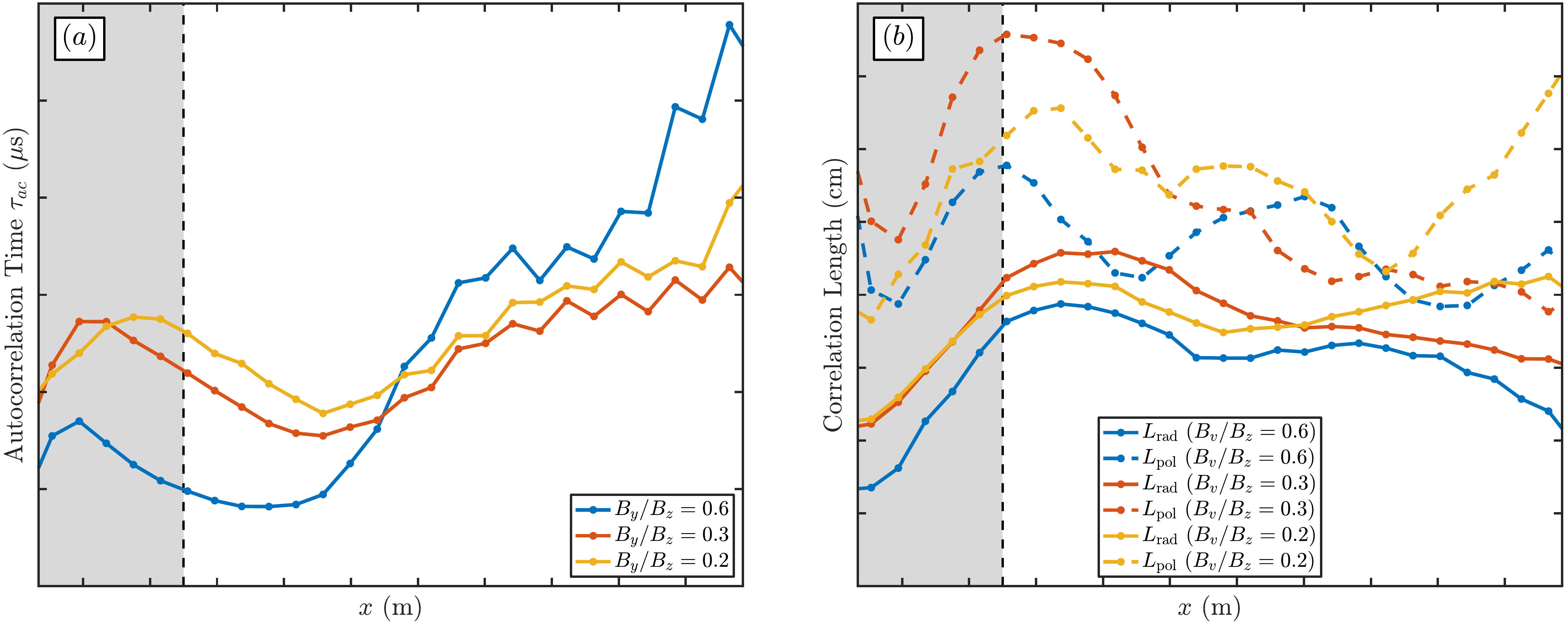}
  \caption[Radial profiles of the autocorrelation times and correlation lengths
  for cases with different magnetic-field-line pitches.]
  {Radial profiles of the ($a$) autocorrelation time and 
  ($b$) poloidal (dashed lines) and radial (solid lines) correlation lengths computed
  at the $z=0$~m plane for three cases with different
  magnetic-field-line pitches. The shaded area indicates
  the region in which the source is concentrated. $L_{\mathrm{pol}}/L_{\mathrm{rad}} \sim
  1.2$--$1.6$ is observed across the radial domain.}
  \label{fig:correlation_lengths} 
\end{figure*}

Figure~\ref{fig:correlation_lengths}($a$) shows the radial profile of the
autocorrelation time $\tau_{ac}$ (computed from time traces of the density fluctuations).
In the SOL of the simulation, $\tau_{ac}$ tends to increase with radius, which is a trend observed in
to measurements on NSTX (see Fig.~12 of \citet{Zweben2015}).
The autocorrelation time for the $B_v/B_z = 0.2$ and $B_v/B_z=0.3$ cases is found to
vary between ${\sim}5$~$\mu$s and ${\sim}9$~$\mu$s, while the autocorrelation time
for the $B_v/B_z = 0.6$ case exhibits a larger variation in the SOL, 
with $\tau_{ac} \approx 4$~$\mu$s for $x < 1.34$~m and increasing to ${\approx}12$~$\mu$s at the
outer radial boundary.
The autocorrelation times we observe in our simulations are lower than the
$\tau_{ac} \sim 10$--$40$~$\mu$s reported by \citet{Zweben2015} for the NSTX edge and SOL,
but are well within the $\tau_{ac} \sim 2$--$20$~$\mu$s range that is typical
for edge and SOL turbulence in other tokamaks.\citep{Boedo2009,Zweben2007}

Figure~\ref{fig:correlation_lengths}($b$) shows the poloidal and radial correlation lengths ($L_{\mathrm{pol}}$ and
$L_{\mathrm{rad}}$ respectively) using the electron-density fluctuations near the $z=0$~m plane.
The correlation length at a radial location is obtained by averaging the correlation length computed at
several points in $y$.
At an individual point, the correlation length is determined from the correlation function, which is
constructed by computing the equal-time two-point autocorrelation function for density fluctuations separated
by some distance $\Delta y$ for $L_{\mathrm{pol}}$ or $\Delta x$ for $L_{\mathrm{rad}}$.
Having observed a significant wave feature in the poloidal correlation function,
we determined $L_{\mathrm{pol}}$ by fitting the poloidal correlation function to
$e^{-|\Delta y|/L_{\mathrm{pol}}} \cos( k_{\mathrm{wave} } \Delta y  )$.
The radial correlation function, which does not have a wave feature,
is computed using the full width at half maximum (FWHM) as $L_{\mathrm{rad}} = \mathrm{FWHM}/(2 \ln 2)$.

For all three values of $B_v/B_z$, we observe that the ratio
$L_{\mathrm{pol}}/L_{\mathrm{rad}}$ is between
1.2 and 1.6 for most of the radial domain, which is similar to the
$L_{\mathrm{pol}}/L_{\mathrm{rad}} \sim 1$--$2$ that is typically observed in tokamaks and stellarators.
\citep{Zweben2007,Boedo2009}
An average $L_{\mathrm{pol}}/L_{\mathrm{rad}} = 1.5 \pm 0.1$ was reported for
representative Ohmic NSTX discharges,\citep{Zweben2016} although larger ratios
$L_{\mathrm{pol}}/L_{\mathrm{rad}} \sim 3$--$4$
have been observed in some experiments \citep{Huber2005} and simulations.\citep{Churchill2017}

\begin{figure}
  \centering
  \includegraphics[width=\linewidth]{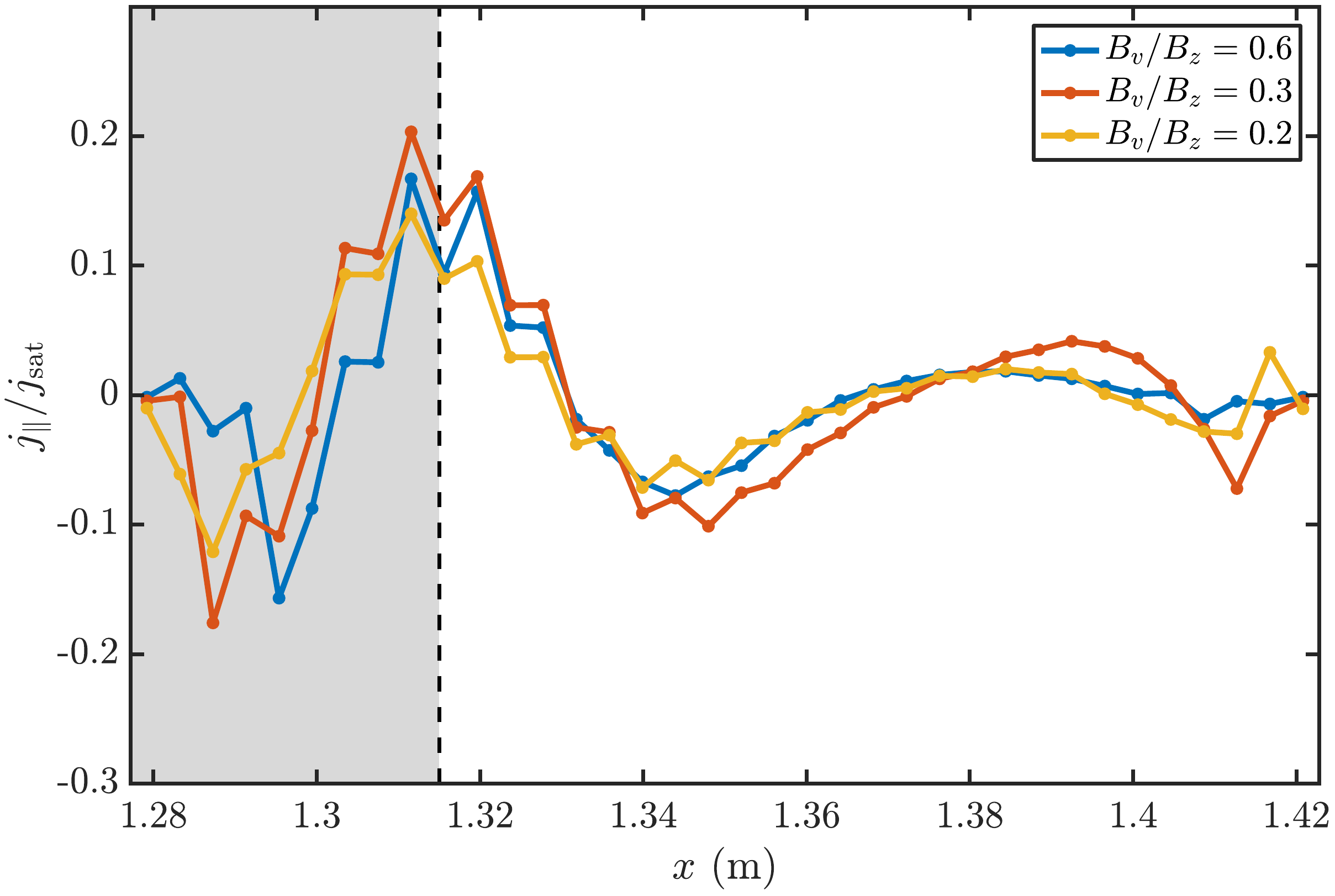}
  \caption{Radial profiles of the steady-state parallel currents into the sheaths for cases
  with different magnetic-field-line pitches.
  The current is normalized to the peak value of the steady-state ion saturation current
  $j_\mathrm{sat} = q_i n_i c_\mathrm{s}$ for each simulation.
  All three cases are quite quantitatively similar, featuring a large excess
  electron outflow in the source region that is balanced by a large excess
  ion outflow just outside of the source region.}
  \label{fig:helical-sol-edge-currents} 
\end{figure}

There are two kinds of sheath-model boundary conditions that are commonly used in fluid and gyrokinetic codes.
Logical-sheath boundary conditions enforce $j_\parallel = 0$ at the sheath entrances, while
current fluctuations into the sheath are permitted in conducting-sheath boundary conditions.
Figure~\ref{fig:helical-sol-edge-currents} shows the radial profiles of the steady-state
parallel current into the sheath for the three cases under consideration.
The currents have been normalized to peak steady-state 
ion saturation current $j_\mathrm{sat} = q_i n_i c_\mathrm{s}$,
where $c_\mathrm{s} = \sqrt{(T_e + \gamma T_i )/m_i}$ and $\gamma = 3$ is used because the collisionless
layer in front of the sheaths should be resolved in all three cases.
All three cases are quite quantitatively similar, and the
outward sheath currents are found to be highly symmetric in $z$, which is consistent
with the strong adiabatic response shown in Fig.~\ref{fig:helical_cross_coherence}.
A large excess electron outflow (negative current) is seen in
the hot source region (near $x=1.3$~m),
which is compensated by a large excess ion outflow (positive current) just
outside the source region.
The peak values are approximately 20\% of the ion saturation current,
which motivates future studies regarding how the use of various sheath-model
boundary conditions affect turbulence in these simulations.

\section{Conclusions \label{sec:conclusions}}
We have developed a model to investigate curvature-driven SOL turbulence in a simplified
helical-magnetic-field geometry and performed numerical simulations of the system using an
electrostatic gyrokinetic continuum code.
The blobs in our simulations appear to originate as radially elongated structures
that extend from the source region into the SOL and get broken up by sheared poloidal flows.
The blobs appear to efficiently transport plasma across the magnetic field, leading to
radial particle fluxes that are much higher than Bohm-flux estimates.
Such large-amplitude and large-scale blobs were not observed in a set of simulations
we performed without magnetic-curvature effects.
We note, however, that coherent structures with high plasma density have been
observed in linear devices with negligible magnetic curvature.\citep{Antar2001,Carter2006}
The mechanism that polarizes such coherent structures in linear devices and leads to
outward radial propagation could be due to neutral wind.\citep{Krasheninnikov2003}

We characterized the turbulence using a variety of diagnostics and found that
various quantities of interest are within the range expected for SOL turbulence in tokamaks,
such as fluctuation levels, autocorrelation times, and correlation lengths.
A summary of some quantities observed in our simulations is given in Table~\ref{tab:helical-experiment},
which also includes experimental values from the NSTX SOL.\citep{Zweben2015,Boedo2014}
We know that there are a number of important physical effects (e.g. complete
magnetic geometry, magnetic fluctuations, and atomic physics) that need to be
added to the simulations in order to expect quantitative accuracy for detailed
comparisons with experiments, but it is interesting to see that the present
simulations are already in the right ballpark qualitatively.
This and other recent work indicate the general feasibility of using continuum
codes to simulate gyrokinetic turbulence in the edge and SOL regions of
tokamaks.

Even in this simple limit we began to explore a number of physical
processes.
We varied the magnetic-field-line pitch in a set of simulations,
which indicated an increasing level of radial turbulent particle transport with decreasing pitch.
A cross-coherence diagnostic comparing potential fluctuations at the sheaths with those at the midplane 
indicated that all three simulations appeared to fall into a similar
turbulent regime with strongly adiabatic electrons.
The application of this model to investigate turbulence in the Helimak device \citep{Gentle2008,Li2011}
has also been performed and will be reported elsewhere.

\begin{table}
\begin{center}
\caption{Summary of helical-SOL simulation results with comparison
  to experimental values for an H-mode NSTX SOL reported in \citet{Zweben2015}.
  The values of $\Gamma_{n,r}$, $T_e$, and $n_e$ refer to values
  near the LCFS (the location of which is not precisely known in the
    experiments \citep{Zweben2004}).
  Since gas-puff imaging cannot be used to obtain particle fluxes,
    the value of $\Gamma_{n,r}$ for the NSTX case is
  taken from \citet{Boedo2014}.
  The `${\sim}$' symbol is used here to indicate that there can be large variations
  in such quantities between discharges with different parameters.
  Ion temperature measurements in the plasma boundary of NSTX were not available,
  so the value of 1--2 (seen on the AUG and MAST tokamaks\citep{Kocan2011}) is assumed.
  }
  \begin{ruledtabular}
  \begin{tabular}{ccc}
    \textbf{Quantity} & \textbf{Simulation Range} & \textbf{NSTX SOL} \\
    $\tau_{ac}$~($\mu$s) & 4--14 & 15--40  \\
    $L_\mathrm{pol}$~(cm) & 2--4 & 3--5 \\
    $L_\mathrm{rad}$~(cm) & 1--2.5 & 2--3  \\
    $\tilde{n}_\mathrm{rms}/\bar{n}$~(\%)& 10--30 & 20--100 \\
    $\Gamma_{n,r}$~$\left(10^{21}~\mathrm{m}^{-2}~\mathrm{s}^{-1} \right)$ & 3.5--5.1 & ${\sim}4$ \\
    $n_e$~$\left(10^{19}~\mathrm{cm}^{-3}\right)$ & 0.5--1.5 & ${\sim}1$ \\
    $T_e$~(eV) & 26--29 & ${\sim}29$ \\
    $T_i/T_e$ & 1.5--2 & 1--2
  \end{tabular}
  \end{ruledtabular}
  \label{tab:helical-experiment}
\end{center}
\end{table}

The helical-SOL model can be extended by the addition of a closed-magnetic-field-line region (with periodic
boundary conditions in the parallel direction).
While the Gkeyll code can already perform simulations with periodicity in the parallel direction,
additional work is required to simultaneously include both open and closed-magnetic-field-line regions
in the same simulation.
The addition of good-magnetic-curvature regions and electromagnetic effects are
also important extensions that will make this model more applicable to
tokamaks.
Since our model is relatively simple compared to a realistic tokamak SOL,
the helical-SOL model could also eventually serve as a test case for the cross verification
of gyrokinetic boundary-plasma codes.
This test case might be useful for revealing major discrepancies due to different
numerical approaches, sheath-model boundary conditions, and collision operators
implemented in various codes relatively early on in the development cycle
before more significant investments are made.

\begin{acknowledgments}
We thank S.\,Zweben for useful discussions about NSTX SOL measurements and J.\,Juno for
setting Gkeyll up on the Stampede cluster.
E.\,L.\,S. would also like to acknowledge useful discussions with J.\,Nichols concerning
ion temperature measurements in the SOL and S.\,Zweben and M.\,Kunz for providing feedback on the manuscript.
This work was funded by the U.S. Department of Energy under Contract DE-AC02-09CH11466, through
the Max-Planck/Princeton Center for Plasma Physics and the Princeton Plasma Physics Laboratory.
E.\,L.\,S. prepared this manuscript in part under the auspices of the U.S. Department of Energy by
Lawrence Livermore National Laboratory under Contract DE-AC52-07NA27344.
G.\,W.\,H. and A.\,H. were supported in part by the SciDAC Partnership for
    Multiscale Gyrokinetic Turbulence.
A.\,H. was also supported in part by the Laboratory Directed Research and Development program.
Some simulations reported in this paper were performed on the Perseus cluster at the TIGRESS high performance
computer center at Princeton University, which is jointly
supported by the Princeton Institute for Computational Science and Engineering and the Princeton
University Office of Information Technology's Research Computing department.
This work also used the Extreme Science and Engineering Discovery Environment (XSEDE),
which is supported by National Science Foundation grant number ACI-1548562.
\end{acknowledgments}

\appendix

\section{Initial Conditions \label{sec:initial-helical-sol}}
We consider a problem in which a uniform mass source $S_\rho$ and energy source $S_E$
is continuously active in the region $|z| < L_S/2$.
This fluid flows out to perfectly absorbing boundaries at $|z| = L_z/2$.
We treat the plasma as a single fluid with mass density
$\rho \approx n_e m_i$, pressure $p = n_e(T_e + T_i) = 2 n_e T$ (where $T$
is an average of the electron and ion temperatures), and
energy density $(3/2)n_e(T_e+T_i)$, so $S_\rho = m_i S_n$ and
$S_E = 3 T_{\mathrm{src}} S_n$,
where $S_n$ is the electron and ion particle source rate and
$T_\mathrm{src} = (T_{e,\mathrm{src}} + T_{i,\mathrm{src}})/2$
is the effective single-fluid source temperature.
This system is described by the steady-state ideal fluid equations
(neglecting thermal conduction and viscosity)
\begin{eqnarray}
  0 &=& -\frac{\partial}{\partial z} \left(\rho u \right) + S_\rho, \\
  0 &=& -\frac{\partial}{\partial z} \left(\rho u^2 + p \right), \\
  0 &=& -\frac{\partial}{\partial z} \left(\frac{1}{2} \rho u^3 + \frac{5}{2} p u \right)  + S_E,
\end{eqnarray}
where $u$ is the fluid velocity, $\rho$ is the mass density, and $p$ is the pressure.
We treat the source as having no mean flow in the $z$ direction.

We integrate these equations from $z=0$ to an arbitrary position $z < L_S/2$
and use the boundary condition $u(z=0) = 0$ to get
\begin{eqnarray}
  \rho u &=& S_\rho z, \\
  \rho u^2 + p &=& p_0, \\
  \frac{1}{2} \rho u^3 + \frac{5}{2} p u &=& z S_E,
\end{eqnarray}
where $p_0 \equiv p(z=0)$.
The first two equations can be solved for $\rho$ and $p$ respectively, and we obtain
a quadratic equation for $u(z)$ by substituting these expressions into the last equation.
The solution to this system is
\begin{eqnarray}
  p(z) &=& \frac{3 p_0 \mp \sqrt{25 p_0^2 - 32 z^2 S_\rho S_E}}{8}, \label{eq:1d_fluid_pressure} \\
  u(z) &=& \frac{5 p_0 \pm \sqrt{25 p_0^2 - 32 z^2 S_\rho S_E}}{8 S_\rho z}, \\
  \rho(z) &=& \frac{z S_p}{u}.
\end{eqnarray}
Since the pressure cannot be negative, the only physical solution for small $z$
is the negative branch for $u(z)$ and the positive branch for $p(z)$.
The central pressure $p_0$ is determined by the boundary conditions at $|z| = L_z/2$.
A steady-state solution at a perfectly absorbing wall requires
$\mathcal{M} \ge 1$ at the wall,\citep{Munz1994} where the Mach number 
$\mathcal{M}(z) \equiv u(z)/c_\mathrm{s}(z) = u(z)/\sqrt{(5/3) p(z) / \rho(z)}$.
The $\mathcal{M} \ge 1$ requirement is equivalent to the Bohm criterion for a
steady-state sheath.
We see that
\begin{equation}
  \mathcal{M}(z)^2 = \frac{\rho(z) u(z)^2}{(5/3) p(z)^2} = \frac{3}{5} \frac{p_0 - p}{p}.
\end{equation}
The maximum possible value of $\mathcal{M}(z)$ occurs at the $z$
that minimizes $p(z)$.
This value $z_\mathrm{max}$ turns out to be the $z$ that makes the
radicand in Eq.~(\ref{eq:1d_fluid_pressure}) zero, so we find that 
\begin{eqnarray}
    z_\mathrm{max}^2 &=& \frac{25}{32} \frac{p_0^2}{S_p S_E},
        \label{eq:zmax} \\
  p(z_\mathrm{max}) &=& \frac{3}{8} p_0, \\
  \mathcal{M}(z_\mathrm{max}) &=& 1.
\end{eqnarray}
This says that the largest possible value of $\mathcal{M}$ is 1
[when $z$ has been made as large as possible, as given by
Eq.~(\ref{eq:zmax})].
This (barely) satisfies the outflow requirement that $\mathcal{M} \ge 1$ at a
perfectly absorbing wall. Equation~(\ref{eq:zmax}) then provides a 
constraint on
the value of $p_0$ such that $\mathcal{M}=1$ is achieved at the end of the
source region, $z_\mathrm{max} = L_S/2$:
\begin{equation}
  p_0 = \frac{L_S}{2} \sqrt{ \frac{32}{25} S_\rho S_E}.
\end{equation}
Using this expression for $p_0$, we have the following profiles in the source
region $0 < |z| < L_S/2$:
\begin{eqnarray}
  p(z) &=& p_0 \left( \frac{ 3 + 5 \sqrt{1-z^2 / \left( L_S/2 \right)^2} }{8} \right), \\
  u(z) &=& \frac{\sqrt{3}}{2} \sqrt{ \frac{ 2 T_\mathrm{src}}{m_i} }
   \left( \frac{ 1 - \sqrt{1-z^2 / \left( L_S/2 \right)^2} }{ z/\left( L_S/2 \right) } \right), \label{eq:1d_fluid_u} \\
  \rho(z) &=& \frac{16 S_\rho^2}{5 p_0} \left( \frac{L_S}{2} \right)^2
    \left( \frac{1 + \sqrt{1-z^2/\left( L_S/2 \right)^2 } }{2} \right).
\end{eqnarray}
In order to use these profiles to initialize a Maxwellian initial condition
for a kinetic simulation, we note that these profiles correspond to density 
($n=\rho/m_i$) and temperature
($T = m_i p/(2 \rho)$) profiles in the source region given by
\begin{eqnarray}
  T(z) &=& \frac{3}{5} T_\mathrm{src} \left(
    \frac{3 + 5 \sqrt{1-z^2 / \left( L_S/2 \right)^2} }
    {4 + 4 \sqrt{1-z^2 / \left( L_S/2 \right)^2} }
    \right), \label{eq:1d_fluid_temp} \\
  n(z) &=& \frac{4\sqrt{5}}{3} \frac{(L_S/2) S_n}{c_\mathrm{ss}} 
    \left( \frac{1 + \sqrt{1-z^2/\left( L_S/2 \right)^2 } }{2} \right), \label{eq:1d_fluid_density}
\end{eqnarray}
where $c_\mathrm{ss} = \sqrt{(5/3) 2 T_\mathrm{src}/m_i}$.
In the source-free regions $z>L_S/2$ or $z<L_S/2$, $n(z)$, $T(z)$, and $u(z)$ are all constant and equal to
the value that their respective profiles evaluated at the corresponding edge of the source region
at $z=L_S/2$ or $z=-L_S/2$.
The 1-D equilibrium profiles Eqs.~(\ref{eq:1d_fluid_u}), (\ref{eq:1d_fluid_temp}), and (\ref{eq:1d_fluid_density}),
the density source in the helical-SOL simulations Eq.~(\ref{eq:helical_sol_source}),
and the temperature profiles of the electron and ion sources are used to
generate spatially varying initial conditions in $(x,y,z)$.

One could go further by calculating the slight difference between the 
ion-guiding-center-density and electron-density profiles
that gives the desired equilibrium potential $\phi(x,y,z)$ when the 
gyrokinetic Poisson equation is solved.
For now, we simply set $n_i^g(x,y,z) = n_e(x,y,z)$ and initialize 
with $\phi = 0$, as we did in the LAPD simulations of \citet{Shi2017}.

\bibliography{shi-helical-sol-2018.bib}

\end{document}